\documentstyle{aipproc2}
\begin{document}
\title{\vspace*{1in}The Challenge Of Data Analysis For Future CMB Observations}

\author{Julian Borrill} 

\address{Center For Particle Astrophysics\thanks{The COMBAT
collaboration is supported by NASA AISRP Grant NAG5-3941.}, University
of California at Berkeley, \& \\ National Energy Research Scientific
Computing Center\thanks{NERSC is supported by the Office of Science of
the U.S. D.o.E under Contract No. DE-AC03-76SF00098}, Lawrence
Berkeley National Laboratory,\\Berkeley, California 94720}

\maketitle

\begin{abstract}
Ongoing observations of the Cosmic Microwave Background, such as the
MAXIMA and BOOMERanG projects, are providing datasets of unprecedented
quality and ever-increasing size. Exact analysis of the data they
produce is a serious computational challenge, currently scaling as the
number of sky pixels squared in memory and cubed in time. Here we
discuss the origins of these scaling relations and their implications
for our efforts to extract precise cosmological parameters from
observations of the CMB.
\end{abstract}

\section*{Introduction}

The Cosmic Microwave Background is the most distant observation of
photons we can ever make. Last scattered only 300,000 years after the
Big Bang it provides a unique picture of the state of the universe at
that time. In particular, fluctuations in the CMB directly trace the
primordial density perturbations and so provide a powerful
discriminant between alternative cosmologies of the very early
universe. As a result the search for anisotropies in the CMB has been
a cornerstone of cosmology for the last 30 years.

Finally measured by the COBE satellite, the anisotropies proved to be
of the order of only one part in a million on a 3K background whose
uniformity was otherwise only broken by a dipole induced by the
peculiar velocity of the galaxy of the order of one part in a
thousand. Despite the tiny scale of these fluctuations, advances in
detector technologies have enabled us to consider measuring them to
the extraordinary accuracy and resolution necessary to determine the
fundamental parameters of cosmology to better than 1\%
\cite{PARAMETERS}.

Such measurements include those of the MAXIMA and BOOMERanG projects
--- described in detail by Lee {\em et al}, and Masi {\em et al} and
de Bernardis {\em et al} elsewhere in these Proceedings. These
balloon-borne observations have already produced datasets an order of
magnitude larger than their predecessors, and in subsequent flights
will at least double this size. Beyond this, the MAP and PLANCK
satellite missions will yield datasets 1-2 orders of magnitude larger
again. The shear size of these datasets makes their analysis a serious
computational challenge. It is this challenge, and the current status
of our attempts to address it, that are discussed here.

For simplicity we only consider the highly idealised case of
extracting a power spectrum of ${\cal N}_{l}$ multipoles in ${\cal
N}_{b}$ bins from a map of ${\cal N}_{p}$ pixels obtained from a
single time-ordered sequence of ${\cal N}_{t}$ observations of the
sky, the data {\em only} comprising CMB signal and Gaussian noise. In
practice there are many additional sources of non-Gaussian
contamination (in particular both galactic and extra-galactic
foreground sources) making observations necessary at a range of
frequencies to allow for their subtraction.

\section*{From The Time-Ordered Data To The Map}

\subsection*{Formalism}

Our first step is to translate the observation from the temporal to
the spatial domain --- to make a map \cite{MAP} (see also Jaffe {\em
et al} elsewhere in these Proceedings). Knowing where the detector was
pointing at each observation, $(\theta_{t}, \psi_{t})$, and adopting a
particular pixelization of the sky, we can construct a pointing matrix
$A_{t p}$ whose entries give the weight of pixel $p$ in observation
$t$. For scanning experiments such as MAXIMA and BOOMERanG this has a
particularly simple form
\begin{equation}
A_{t p} = \left\{ \begin{array}{ll}
	1 & {\rm if} \;\;\; (\theta_{t}, \psi_{t}) \in p \\
	0 & {\rm otherwise}
	\end{array}
\right.
\end{equation}
while other observing strategies would give a more complex
structure. The data vector can now be written
\begin{equation}
\label{eDSN}
d = A s + n
\end{equation}
in terms of the pixelised CMB signal $s$ and time-stream noise $n$. 

Under the assumption of Gaussianity, the noise probability
distribution is
\begin{equation}
\label{eTTNPD}
P(n) = (2 \pi)^{-{\cal N}_{t}/2} \exp \left\{ -\frac{1}{2} \left(
n^{T} N^{-1} n + Tr \left[ \ln N \right] \right) \right\}
\end{equation}
where ${\cal N}$ is the time-time noise correlation matrix given by
\begin{equation}
N \equiv \langle n n^{T} \rangle
\end{equation}
We can now use equation (\ref{eDSN}) to substitute for the noise in
equation (\ref{eTTNPD}), giving the probability of the data for a
particular CMB signal as
\begin{equation}
P(d | s) = (2 \pi)^{-{\cal N}_{t}/2} \exp \left\{ -\frac{1}{2} \left(
(d - A s)^{T} N^{-1} (d - A s) + Tr \left[ \ln N \right] \right)
\right\}
\end{equation}
Assuming that all CMB maps are {\em a priori} equally likely, this is
proportional to the likelihood of the signal given the data, and
maximizing over $s$ gives the maximum likelihood map $m$
\begin{equation}
\label{eMAP}
m = \left( A^{T} N^{-1} A \right)^{-1} A^{T} N^{-1} d 
\end{equation}

Substituting back for the time-ordered data in this map we recover the
obvious fact that it is the sum of the true CMB signal and some
pixelized noise
\begin{eqnarray}
m & = & \left( A^{T} N^{-1} A \right)^{-1} A^{T} N^{-1} (A s + n) \nonumber \\
  & = & s + \nu
\end{eqnarray}
where this pixel noise
\begin{equation}
\nu = \left( A^{T} N^{-1} A \right)^{-1} A^{T} N^{-1} n
\end{equation}
has correlations given by
\begin{eqnarray}
\Upsilon & = & \langle \nu \nu^{T} \rangle \nonumber \\
         & = & \left( A^{T} N^{-1} A \right)^{-1}
\end{eqnarray}

\subsection*{Computational Requirements}

Making the map requires solving equation (\ref{eMAP}) which is
conveniently divided into three steps:
\begin{eqnarray}
\Upsilon^{-1} & = & A^{T} N^{-1} A \nonumber \\
z & = & A^{T} N^{-1} d \nonumber \\
{\rm and} \;\;\;\;\;\;\;\;\;\;\;\;\;\;\;\;\;\;\;\;\;\;\;\;
m & = & (\Upsilon^{-1})^{-1} z
\end{eqnarray}
The first half of table \ref{tMM} shows the computational cost of a
brute force approach to each of these steps (recall that multiplying
an $[a \times b]$ matrix and an $[b \times c]$ matrix involves $2
\times a \times b \times c$ operations). Thus for datasets such as
MAXIMA-1 or BOOMERanG North America, with ${\cal N}_{t} \sim 2 \times
10^{6}$ and ${\cal N}_{p} \sim 3 \times 10^{4}$, making the map would
require of the order of 16 Tb of disc space (storing data in 4-byte
precision), 7 Gb RAM\footnote{Although it is possible to use
out-of-core algorithms for operations such as matrix inversion the
associated time overhead would be prohibative. We therefore assume
that all such operations are carried out in core.} (doing all
calculations in 8-byte precision) and $2.4 \times 10^{17}$ operations.
If we were able to use a fast 600 MHz CPU at 100\% efficiency this
would still correspond to over 12 years of run time.

\begin{table*}[t!]
\caption{Computational requirements for the map-making algorithm}
\label{tMM}
\begin{tabular}{ccccccc}
Calculation & \multicolumn{3}{c}{Brute Force} & \multicolumn{3}{c}{Structure-Exploiting} \\
& Disc & RAM & Operations & Disc & RAM & Operations \\
\tableline
\vspace*{-0.1in}&&&&&\\
\vspace*{0.05in}$N^{-1}_{p p'} = A^{T} N^{-1} A $ & 
$4 {\cal N}_{t}^{2}$ & 
$16 {\cal N}_{t}$ & 
$2 {\cal N}_{p} {\cal N}_{t}^{2}$ & 
$4 {\cal N}_{p}^{2}$ &
$8 {\cal N}_{p}^{2}$ & 
$3 {\cal N}_{\alpha}^{2} {\cal N}_{\tau} {\cal N}_{t}$ \\
\vspace*{0.05in}$z = A^{T} N^{-1} d$ & 
$4 {\cal N}_{t}^{2}$ & 
$16 {\cal N}_{t}$ & 
$2 {\cal N}_{t}^{2}$ & 
$4 {\cal N}_{t}$ & 
$8 {\cal N}_{t}$ & 
$3 {\cal N}_{\alpha} {\cal N}_{\tau} {\cal N}_{t}$ \\
\vspace*{0.05in}$m = (\Upsilon^{-1})^{-1} z$ & 
$4 {\cal N}_{p}^{2}$ & 
$8 {\cal N}_{p}^{2}$ & 
$(2 + \frac{2}{3}) {\cal N}_{p}^{3}$ & 
$4 {\cal N}_{p}^{2}$ & 
$8 {\cal N}_{p}^{2}$ & 
$(2 + \frac{2}{3}) {\cal N}_{p}^{3}$ \\
\end{tabular}
\end{table*}

Fortunately there are two crucial structural features to be exploited
here. As noted above the pointing matrix $A$ is usually very sparse,
with only ${\cal N}_{\alpha}$ non-zero entries in each row. For simple
scanning strategies such as MAXIMA, BOOMERanG and PLANCK, ${\cal
N}_{\alpha} = 1$, with a single 1 in the column corresponding to the
pixel being observed. For a differencing experiment such as COBE or
MAP, ${\cal N}_{\alpha} = 2$, with a $\pm 1$ pair in the columns
corresponding to the pixel pair being observed. Moreover, the inverse
time-time noise correlations are (by fiat) both stationary and fall to
zero beyond some time-separation much shorter than the duration of the
observation
\begin{eqnarray}
N^{-1}_{t t'} & = & f(|t - t'|) \nonumber \\
              & = & 0 \;\;\;\;\;\; \forall \;\;\; |t - t'| > \tau \ll {\cal N}_{t}
\end{eqnarray}
so that the inverse time-time noise correlation matrix is symmetric
and band-diagonal, with bandwith ${\cal N}_{\tau} = 2 \tau + 1$. The second
half of table \ref{tMM} shows the impact of exploiting this structure
on the cost of each step. The limiting step is now no longer
constructing the inverse pixel-pixel noise correlation matrix but
solving for the map, which is unaffected by these features. For the
same datasets making the map now takes of the order of 3.6 Gb of disc,
7 Gb of RAM, and $7 \times 10^{13}$ operations, or 32 hours of the
same CPU time.

Further acceleration of the map-making algorithm must therefore focus
on a faster solution the final step, inverting the inverse pixel-pixel
noise covariance matrix $\Upsilon^{-1}$ to obtain the map. However, as
we shall see below, even this is not the limiting step overall in
current algorithms.

\section*{From The Map To The Power Spectrum}

\subsection*{Formalism}

We now want to move to a realm where the CMB observation can be
compared with the predictions of various cosmological theories ---
typically the angular power spectrum. We decompose the CMB signal at
each pixel in spherical harmonics
\begin{equation}
s_{p} = \sum_{l m} a_{l m} B_{l} Y_{l m}(\theta_p, \psi_p)
\end{equation}
where $B$ is the pattern of the observation beam (assumed to be
circularly symmetric) in $l$-space. The correlations between such
signals then become
\begin{equation}
S_{p p'} \equiv \langle s_{p} s_{p'} \rangle = \sum_{l m} \sum_{l' m'}
\langle a_{l m} a_{l' m'} \rangle B_{l} B_{l'} Y_{l m}(\theta_p, \psi_p) Y_{l'
m'}(\theta_{p'}, \psi_{p'})
\end{equation}
For isotropic fluctuations the correlations depend only on the angular
separation
\begin{equation}
\langle a_{l m} a_{l' m'} \rangle = C_{l} \delta_{l l'} \delta_{m m'}
\end{equation}
and the pixel-pixel signal correlation matrix becomes
\begin{equation}
S_{p p'} = \sum_{l} \frac{2 l + 1}{4 \pi} C_{l} P_{l}(\chi_{p p'})
\end{equation}
where $P_{l}$ is the Legendre polynomial and $\chi_{p p'}$ the angle
between the pixel pair $p, p'$. These $C_{l}$ multipole powers completely
characterise a Gaussian CMB, and are an otherwise model-independent
basis in which to compare theory with observations. In general, due to
incomplete sky coverage and low signal-to-noise, we are unable to
extract each multipole moment independently. We therefore group the
multipoles in bins, adopting a particular spectral shape function
$C_{l}^{s}$ and characterising the CMB signal by its bin powers
$C_{b}$ with
\begin{equation}
C_{l} = C_{b: l \in b} C_{l}^{s}
\end{equation}

Since the signal and noise are assumed to be realisations of independent Gaussian
processes the pixel-pixel map correlations are 
\begin{eqnarray}
M_{p p'} & \equiv & \langle m m_{T} \rangle \nonumber \\
	 & = & \langle s s_{T} \rangle + \langle \nu \nu^{T} \rangle \nonumber \\
	 & = & S + \Upsilon
\end{eqnarray}
and the probability distribution of the map given a particular power
spectrum ${\cal C}$ is now
\begin{equation}
\label{eMLA}
P(m | {\cal C}) = (2 \pi)^{-{\cal N}_{p}/2} \exp \left\{
-\frac{1}{2} \left( m^{T} M^{-1} m + Tr \left[ \ln M \right] \right) \right\}
\end{equation}
Assuming a uniform prioir for the spectra, this is proportional to the
likelihood of the power spectrum given the map. Maximizing this over
${\cal C}$ then gives us the required result, namely the most likely
CMB power spectrum underlying the original observation $d$.

Finding the maximum of the likelihood function of equation
(\ref{eMLA}) is generically a much harder problem than making the
map. Since there is no closed-form solution corresponding to equation
(\ref{eMAP}) we must find both a fast way to evaluate the likelihood
function at a point, and an efficient way to search the ${\cal
N}_{b}$-dimensional parameter space for the peak. The fastest general
method extant is to use Newton-Raphson iteration to find the zero of
the derivative of the logarithm of the likelihood function
\cite{MLA}. If the log likelihood function
\begin{equation}
{\cal L}({\cal C}) = - \frac{1}{2} \left( m^{T} M^{-1} m + Tr \left[ \ln M \right] \right)
\end{equation}
were quadratic, then starting from some initial guess at the maximum
likelihood power spectrum ${\cal C}_{o}$ the correction $\delta {\cal
C}_{o}$ that would take us to the true peak would simply be
\begin{equation}
\label{eDC}
\delta {\cal C}_{o} = - \left( \left[ \frac{\partial^2 {\cal L}}{\partial {\cal C}^{2}} \right]^{-1} 
\frac{\partial {\cal L}}{\partial {\cal C}} \right)_{{\cal C} = {\cal C}_{o}}
\end{equation}
Since the log likelihood function is not quadratic, we now take
\begin{equation}
{\cal C}_{1} = {\cal C}_{o} + \delta {\cal C}_{o}
\end{equation}
and iterate until $\delta {\cal C}_{n} \sim 0$ to the desired
accuracy. Because any function is approximately quadratic near a peak,
if we start searching sufficiently close to a peak this algorithm will
converge to it. Of course there is no guarantee that it will be the
global maximum, and in general there is no certainty about what
`sufficiently close' means in practice. However experience to date
suggests that the log likelihood function is sufficiently strongly
singley peaked to allow us to use this algorithm with some confidence.

\subsection*{Computational Requirements}

The core of the algorithm is then to calculate the first two
derivatives of the log likelihood function with respect to the
multipole bin powers
\begin{eqnarray}
\frac{\partial {\cal L}}{\partial {\cal C}_{b}} & = & 
\frac{1}{2} \left( m^{T} M^{-1} \frac{\partial S}{\partial {\cal C}_{b}} M^{-1} m
- Tr \left[ M^{-1} \frac{\partial S}{\partial {\cal C}_{b}} \right] \right) \\ 
\frac{\partial^2 {\cal L}}{\partial {\cal C}_{b} \partial {\cal C}_{b'}} & = & 
- m^{T} M^{-1} \frac{\partial S}{\partial {\cal C}_{b}} M^{-1} 
\frac{\partial S}{\partial {\cal C}_{b'}} M^{-1} m +
\frac{1}{2} Tr \left[ M^{-1} \frac{\partial S}{\partial {\cal C}_{b}} M^{-1}
\frac{\partial S}{\partial {\cal C}_{b'}}\right]
\end{eqnarray}
Evaluating these derivatives comes down to solving the ${\cal N}_{b}
{\cal N}_{p} + 1$ linear systems
\begin{eqnarray}
z & = & M^{-1} m \\
{\rm and} \;\;\;\;\;\;\;\;\;\;\;\;\;\;\;\;\;\;\;\;\;\;\;\; 
W_{b} & = & M^{-1} \frac{\partial S}{\partial {\cal C}_{b}} \;\;\; \forall \;\;\; b
\end{eqnarray}
and assembling the results
\begin{eqnarray}
\frac{\partial {\cal L}}{\partial {\cal C}_{b}} & = & 
\frac{1}{2} \left( m^{T} W_{b} z - Tr \left[ W_{b} \right] \right) \nonumber \\
\frac{\partial^2 {\cal L}}{\partial {\cal C}_{b} \partial {\cal C}_{b'}} & = & 
- m^{T} W_{b} W_{b'} z + \frac{1}{2} Tr \left[ W_{b} W_{b'} \right]
\end{eqnarray}
Table \ref{tMLA} shows the computational cost of these operations,
where solving the linear systems has been optimised by first Cholesky
decomposing the matrix $M$. Solving equation (\ref{eDC}) has been
excluded since its cost is entirely negligible, depending only on the
number of multipole bins ${\cal N}_{b} \ll {\cal N}_{p}$. Obtaining
the maximum likelihood power spectrum for the same datasets as above,
with ${\cal N}_{p} \sim 3 \times 10^{4}$ and ${\cal N}_{b} \sim 20$,
then requires of the order of 150 Gb disc, 14 Gb of RAM, and $10^{15}$
operations per iteration, or 21 days of our 600 MHz CPU time.

\begin{table*}[t!]
\caption{Computational requirements for each iteration of the maximum
likelihood power spectrum extraction algorithm}
\label{tMLA}
\begin{tabular}{cccc}
Calculation & Disc & RAM & Operations \\
\tableline
\vspace*{-0.1in}&&&\\
\vspace*{0.05in} $M  = L L^{T}$ & 
$4 {\cal N}_{p}^{2}$ & 
$8 {\cal N}_{p}^{2}$ & 
$\frac{2}{3} {\cal N}_{p}^{3}$ \\
\vspace*{0.05in} $L L^{T} z = m$ &
$4 {\cal N}_{p}^{2}$ & 
$8 {\cal N}_{p}^{2}$ & 
$2 {\cal N}_{p}^{3}$ \\
\vspace*{0.05in} $L L^{T} W_{b} = \frac{\partial S}{\partial {\cal C}_{b}}$ &
$8 {\cal N}_{b} {\cal N}_{p}^{2}$ & 
$16 {\cal N}_{p}^{2}$ & 
$2 {\cal N}_{b} {\cal N}_{p}^{3}$ \\
\vspace*{0.05in} $\frac{\partial {\cal L}}{\partial {\cal C}_{b}} = 
\frac{1}{2} \left( m^{T} W_{b} z - Tr \left[ W_{b} \right] \right)$ &
$4 {\cal N}_{b} {\cal N}_{p}^{2}$ & 
$8 {\cal N}_{p}^{2}$ & 
$2 {\cal N}_{b} {\cal N}_{p}^{2}$ \\
\vspace*{0.05in} $\frac{\partial^2 {\cal L}}{\partial {\cal C}_{b} \partial {\cal C}_{b'}} = 
- m^{T} W_{b} W_{b'} z + \frac{1}{2} Tr \left[ W_{b} W_{b'} \right]$ \hspace*{0.25in} &
$4 {\cal N}_{b} {\cal N}_{p}^{2}$ & 
$16 {\cal N}_{p}^{2}$ & 
$3 {\cal N}_{b}^{2} {\cal N}_{p}^{2}$ \\
\end{tabular}
\end{table*}

Such numbers are at least conceivable; however, as shown in table
\ref{tCRR}, the scaling with map size pushes forthcoming balloon
observations well beyond the capacity of the most powerful single
processor machine --- and even allowing for the continued doubling of
computer power every 18 months predicted by Moore's `law' we would
still have to wait 20 years for a serial machine capable of handling
the BOOMERanG Long Duration Balloon flight data. Moreover, these
timings are for a single iteration (and typically the alogorithm needs
${\cal O}(10)$ iterations to converge) for a single run through the
dataset, although undoubtedly we will want to perform several slightly
different runs to check the robustness of our analysis.

One way to increase our capability now is to move to parallel
machines, such as the 640-processor Cray T3E at NERSC. Since the
algorithm is limited by matrix-matrix operations (Cholesky
decomposition and triangular solves) we are able to exploit the most
optimised protocols --- the level 3 BLAS --- and the DEC Alpha chips'
capacity to perform an add and a multiply in a single clock
cycle. Coupled with a finely-tuned dense packing of the matrices on
the processors this has enabled us to sustain upwards of 2/3 of the
theoretical peak performance of 900MHz. This enables us to increase
the limiting datasize to around 80,000 pixels.

\begin{table*}[t!]
\caption{The computational requirements for one iteration of the
Newton-Raphson algorithm to extract a 20-bin power spectrum for MAXIMA
and BOOMERanG}
\label{tCRR}
\begin{tabular}{ccccccc}
Flight & ${\cal N}_{p}$ & Disc & RAM & Operations & Serial Time & Cray T3E Time\\ 
\tableline 
\vspace*{-0.1in}&&&&&\\
\vspace*{0.05in}BOOMERanG NA  &  26,000 & 110 Gb &  11 Gb & $7.1 \times 10^{14}$ &   14 days &   5 hours (64 PE) \\ 
\vspace*{0.05in}MAXIMA 1      &  32,000 & 170 Gb &  17 Gb & $1.3 \times 10^{15}$ &   25 days &   9 hours (64 PE) \\ 
\vspace*{0.05in}MAXIMA 2      &  80,000 &   1 Tb & 100 Gb & $2.1 \times 10^{16}$ & 13 months & 18 hours (512 PE) \\
\vspace*{0.05in}BOOMERanG LDB & 450,000 &  30 Tb &   3 Tb & $3.7 \times 10^{18}$ & 196 years & 140 days (512 PE) \\
\end{tabular}
\end{table*}

\section*{Future Prospects}

We have seen that existing algorithms are capable of dealing with CMB
datasets with at most $10^{5}$ pixels. Over the next 10 years a range
of observations are expected to produce datasets of $5 \times 10^{5}$
(BOOMERanG LDB), $10^{6}$ (MAP) and $10^{7}$ (PLANCK) pixels that will
necessarily require new techniques. This is an ongoing area of
research in which some progress has been made in particular special
cases.

The limiting steps in the above analysis are associated with
operations involving the pixel-pixel correlation matrices for the
noise $\Upsilon$, the signal $S$, and most particularly their sum
$M$. The problem here is the noise and the signal have different
natural bases. The inverse noise correlations are symmetric,
band-diagonal and approximately circulant in the time domain, while
the signal correlations are diagonal in the spherical harmonic
domain. However there is no known basis in which the map correlations
take a similarly simple form.

If the noise is uncorrelated between pixels {\em and} is azimuthally
symmetric --- as has been argued will be the case for the MAP
satellite due to its chopped observing strategy --- then the
pixel-pixel data correlation matrix can be dramatically simplified,
reducing the analysis to ${\cal O}({\cal N}_{p}^{3/2})$ in storage and
${\cal O}({\cal N}_{p}^{2})$ in operations \cite{APPROX}. Although
some work has also been done extending this to observations with
marginal azimuthal asymmetry it is still inapplicable for the
spatially correlated noise inherent to the simple scanning strategies
adopted by MAXIMA and BOOMERanG (which also face the problem of
irregular sky coverage) or PLANCK; for such observations the problem
remains unsolved.

\section*{Acknowledgements}

The author would like to thank the organisers of the Rome 3K Cosmology
Euroconference for the opportunity to participate in an excellent
meeting, and Andrew Jaffe, Pedro Ferriera, Shaul Hanany, Xiaoye Li,
Osni Marques, and Radek Stompor for many useful discussions.

\end{document}